\documentclass[]{spieman}  

\usepackage{amsmath,amsfonts,amssymb}
\usepackage{graphicx}
\usepackage[]{hyperref}
\usepackage{aas_macros}
\usepackage{siunitx}

\usepackage{caption}
\usepackage{subcaption}

\usepackage{placeins}

\usepackage[]{threeparttable}

\usepackage{xcolor, soul}
\sethlcolor{yellow}

\usepackage{textcomp}

\newcommand{\aeff}{A\textsubscript{eff}}

\newsavebox{\twosubbox}

\newsavebox{\secsubbox}

\newsavebox{\aeffsubbox}

\title{FLUID: A rocket-borne pathfinder instrument for high efficiency UV band selection imaging}

\author[a, *]{Nicholas Nell}
\author[a]{Nicholas Kruczek}
\author[a]{Kevin France}
\author[a]{Stefan Ulrich}
\author[a]{Patrick Behr}
\author[a]{Emily Farr}

\affil[a]{Laboratory for Atmospheric and Space Physics, 1234 Innovation Drive, Boulder, CO, USA}

\begin{document} 
\maketitle



\begin{abstract}
The Far- and Lyman-Ultraviolet Imaging Demonstrator (FLUID) is a
rocket-borne arcsecond-level ultraviolet (UV) imaging instrument
covering four bands between 92 -- 193 nm. FLUID will observe nearby
galaxies to find and characterize the most massive stars that are the
primary drivers of the chemical and dynamical evolution of galaxies,
and the co-evolution of the surrounding galactic environment. The
FLUID short wave channel is designed to suppress efficiency at
Lyman-$\alpha$ (121.6 \si{\nm}), while enhancing the reflectivity of
shorter wavelengths. Utilizing this technology, FLUID will take the
first ever images of local galaxies isolated in the Lyman ultraviolet
(90 -- 120 \si{\nm}). As a pathfinder instrument, FLUID will employ
and increase the TRL of band-selecting UV coatings, and solar-blind UV
detector technologies including microchannel plate and solid state
detectors; technologies prioritized in the 2022 NASA Astrophysical
Biennial Technology Report. These technologies enable high throughput
and high sensitivity observations in the four co-aligned UV imaging
bands that make up the FLUID instrument. We present the design of
FLUID, status on the technology development, and results from initial
assembly and calibration of the FLUID instrument.
\end{abstract}


{\noindent \footnotesize\textbf{*}Nicholas Nell,  \linkable{nicholas.nell@lasp.colorado.edu} }

\section{INTRODUCTION}
\label{sec:intro}  
\begin{figure}[!ht]
  \centering
  \includegraphics[width=\textwidth]{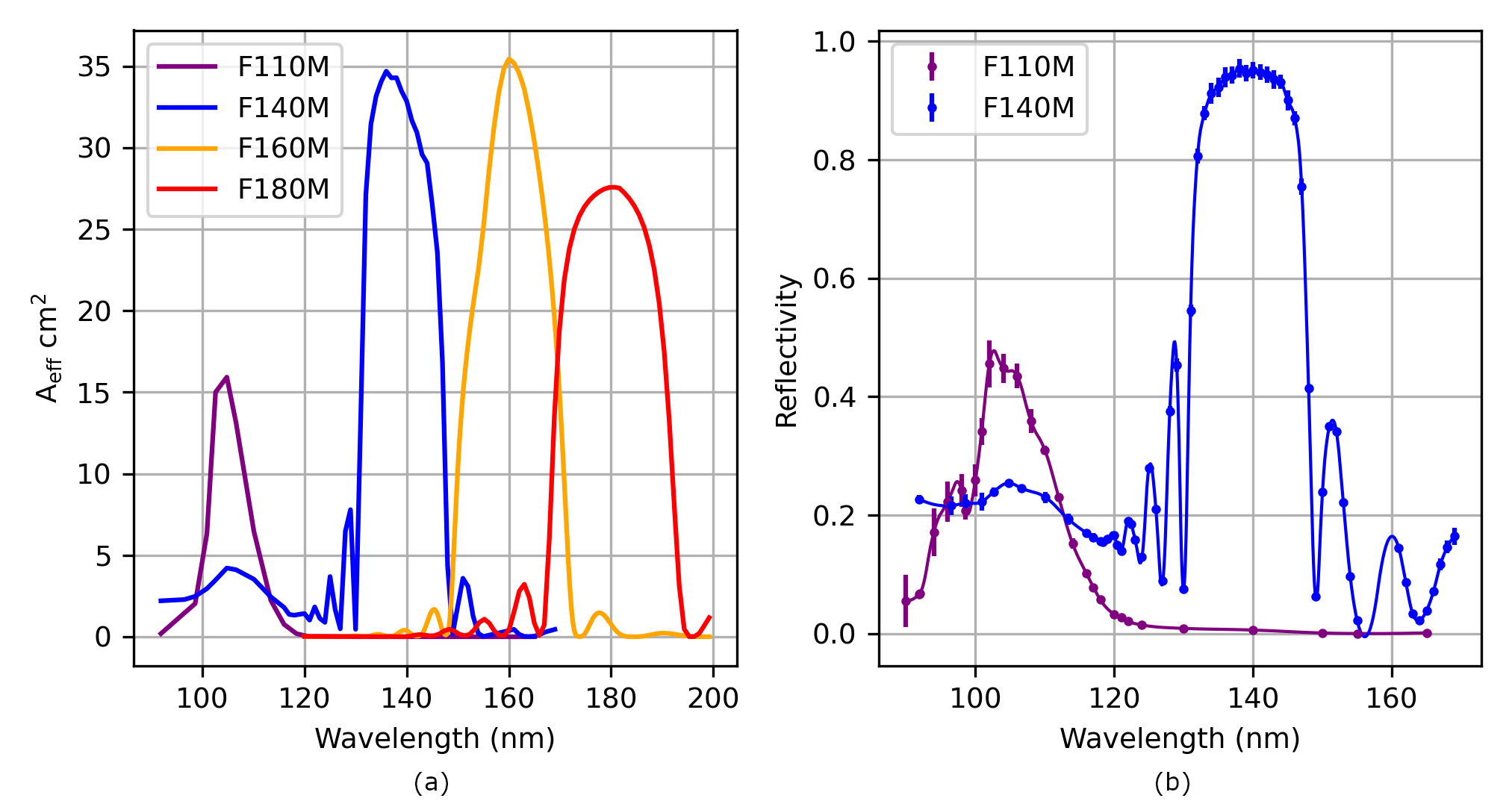}
  \caption{\label{fig:aeff}(a) Predicted effective area for each channel of FLUID
    based on empirical and theoretical coating efficiencies. F110M and
    F140M are based off of empirical reflectivity measurements of
    coated optics, F160M and F180M are based off of theoretical
    coating efficiencies. (b) Empirical measurements performed at LASP
    of primary mirrors for the F110M and F140M channels (see Section
    \ref{sec:fuv_filters}).}
\end{figure}

\begin{figure}[]
  \centering
  \includegraphics[height=0.95\textheight]{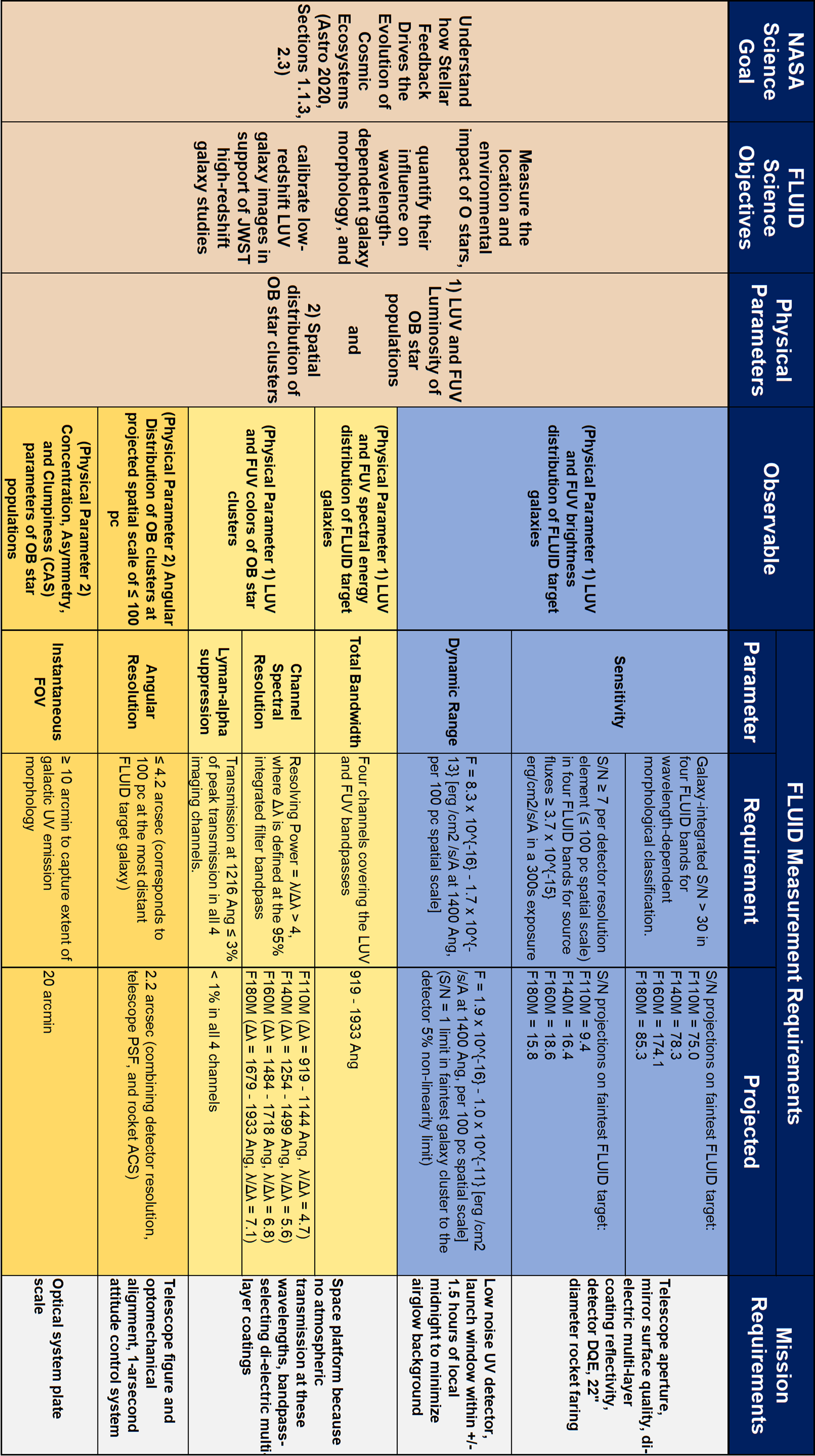}
  \caption{\label{fig:stm}FLUID science traceability matrix.}
\end{figure}

FLUID is a proposed rocket-borne arcsecond-level, multi-band imaging
system operating across the Far UV (FUV; 120 -- 200 nm) and Lyman UV
(LUV; 90 -- 120 nm) bands (see Figure \ref{fig:aeff}), designed to
investigate massive-star formation in nearby galaxies and obtain the
first morphological classification of nearby galaxies in the
LUV\cite{Nell2023}. The highest mass stars (O- and B-type stars;
$\approx$ 5 - 200 M$_{\rm solar}$\cite{Crowther10}), born in
short-lived bursts of star formation, dominate the stellar luminosity
of these early galaxies and are the strongest drivers of a galaxy's
on-going kinematic and chemical
evolution\cite{grimes09,Kobayashi11,Langer12}. The FLUID instrument
will leverage a novel selection of channels to image local galaxies
across the LUV and FUV bandpasses for the first time. The four
channels selected for FLUID offer an observational handle to
concurrently and consistently solve for the stellar temperature and
dust extinction on spatial scales comparable to the size of star
forming regions in those galaxies ($\sim100$ pc), an observation which
is not possible using current orbital assets without costly
spectroscopic observations of every cluster. These data, combined with
archival observations, enable us to quantify the influence of O stars
on the mass-chemical-energy cycle of the target galaxies. FLUID
spatial resolution on local galaxies was further designed to match the
resolution that JWST is currently achieving on high-z
galaxies. Through the use of morphological classification, FLUID
observations will be used in conjunction with JWST observations to
track the spatial distribution of massive stars across cosmic
time. The science traceability matrix (Figure \ref{fig:stm}) is used
to inform and direct the instrument design and technology development
discussed in the following sections. Currently, funding has been
obtained to fabricate two channels of FLUID, F110M and F140M, which
are actively being built and characterized.

\section{INSTRUMENT DESIGN}

\subsection{Instrument Overview}

\begin{figure}[!ht]
  \centering
  \includegraphics[width=\textwidth]{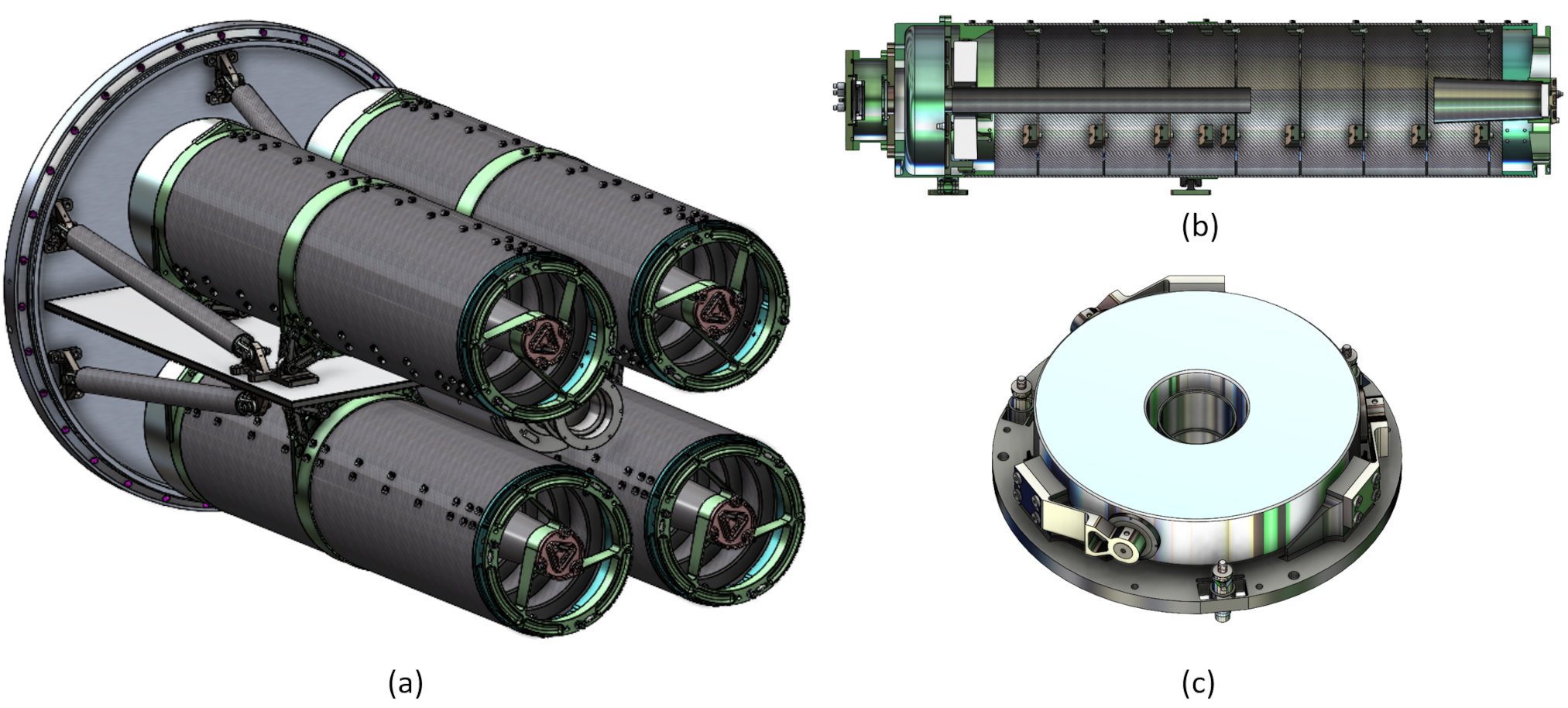}
  \caption{\label{fig:instrument}(a) A CAD rendering of the full FLUID
    instrument. The four telescope modules are mounted to a hermetic,
    22'' bulkhead, and arranged about the roll axis of the rocket. The
    NASA-provided ST5K star tracker\cite{Percival2008} is shown at the
    center of the telescope array. (b) A cutaway view of a single
    FLUID telescope. (c) A CAD rendering of the primary mirror mount.}
\end{figure}

The FLUID optical assembly comprises an array of four imaging
channels, housed within a standard NASA Wallops Flight Facility (WFF)
22'' diameter rocket skin (Figure \ref{fig:instrument}). Each
telescope module includes a 150 mm diameter {\em f}/28.7 Cassegrain
telescope feeding either a CCD or microchannel plate (MCP) detector,
depending on the channel (see Figure \ref{fig:instrument} and Table
\ref{tab:instrument}). The detector-area limited field-of-view has a
\ang{;20;} diameter. The telescope optics for F110M and F140M were
fabricated by a division of Coherent (formerly Precision Asphere) with
polishing capabilities, such as low microroughness and mid-spatial
frequency control and metrology, demonstrated on UV
optics\cite{Kong2014}. Two sets of telescope optics were acquired with
the current funding and the remaining two sets will be purchased
pending future proposal selection. Filters for the FLUID instrument
are all reflective in nature and consist of multilayer coatings
applied to the telescope optics optimized for high reflectivity of
in-band light and low reflectivity of out-of-band light. The FLUID
flight filters are defined by the primary and secondary telescope
mirrors, each coated with a multilayer prescription specific to their
designated bandpass (Section \ref{sec:techdev}). The two-bounce
reflection filter improves the out-of-band rejection, matching the
configuration of {\em LUVOIR}/LUMOS\cite{France2017}, an instrument
which serves as a design prototype for a UV imaging and spectroscopy
instrument for the upcoming {\em Habitable Worlds Observatory}. The
LUMOS design utilizes 150 mm diameter shaped fold mirrors, the same
size as the FLUID primary mirrors. The FLUID primary mirrors therefore
provide an excellent prototype for TRL demonstration. A summary of the
instrument optical specifications and effective areas (\aeff{}) can be
found in Figure \ref{fig:aeff} and Table
\ref{tab:instrument}. Effective areas for F110M and F140M are created
using measured reflectivities of the flight optics and predicted
efficiencies of the detectors. Effective areas for F160M and F180M are
created using theoretical coating efficiencies and predicted detector
efficiencies.


\begin{table}
  \caption{FLUID Instrument Summary}
  \label{tab:instrument}
  \begin{center}
\begin{tabular}{  l   l }
\hline
\multicolumn{2}{c}{Instrument Parameters} \\ \hline
Number of Channels & 4 \\
Focal Ratio & {\em f}/28.7 \\
Bandpass & 91.9 -- 193.3 nm \\
Field of View & \ang{;20;} diameter \\
Instrument Plate Scale & \ang{;;48} mm$^{-1}$ \\
Instrument Peak \aeff{} (F110M) & 16 cm$^2$ \\
Instrument Peak \aeff{} (F140M) & 35 cm$^2$ \\
Instrument Peak \aeff{} (F160M) & 35 cm$^2$ \\
Instrument Peak \aeff{} (F180M) & 28 cm$^2$ \\
Effective Focal Length & 4300.3 mm \\
Total Instrument Length & 870 mm \\ \hline
\multicolumn{2}{c}{Cassegrain Parameters} \\ \hline
Primary Diameter & 150.0 mm \\
Primary Radius of Curvature & 1800.0 mm \\
Secondary Diameter & 42.0 mm \\
Secondary Radius of Curvature & 459.8 mm \\ \hline
\multicolumn{2}{c}{Detector Parameters} \\ \hline
Number of Detectors & 4 \\
F110M & CsI MCP \\
F140M & CsI MCP \\
F160M & Passivated CCD \\
F180M & Bialkali MCP tube \\
\end{tabular}
\end{center}
\end{table}

Each FLUID channel will employ its own bandpass-optimized
detector. The F110M and F140M channel detectors are 25 mm active
diameter open-face cross-delay line (XDL) MCP detectors with opaque
CsI photochathodes applied. The F140M, F160M, and F180M filters
maintain low efficiency ($\sim$5\%) at visible wavelengths allowing us
to utilize detector technologies optimized for a specific UV bandpass
without concern for significant contamination from visible
wavelengths. The F110M filter is reflective at visible wavelengths,
but the photocathode used for this channel, CsI, is not responsive at
these wavelengths\cite{DeMarcos_2018, Tremsin_2001}. F180M will
utilize a 50 mm square sealed tube MCP\cite{Siegmund2023} with a
semi-transparent bialkali photocathode. Not only does this
photochathode selection maximize the efficiency for FLUID science
measurements in each bandpass, it is also the configuration baselined
for {\em LUVOIR}/LUMOS\cite{France2017}.

FLUID will employ a CCD detector to maximize the throughput of the
F160M channel (e.g., a passivated
CCD\cite{Stern1994}). Characterization results have been published for
both surface passivation and anti-reflection (AR) coatings available
from Teledyne e2v that substantially improve quantum efficiency (QE)
in the FUV bandpass\cite{Heymes2020b, Heymes2020a}. The initial flight
of FLUID will have a 27.6 mm square CCD42-40 with enhanced passivation
and the ``VUV1'' AR coating from Teledyne e2v serve as the detector
for the F160M channel. The CCD42-40 was chosen for its compatibility
with the FLUID design and extensive space flight
heritage\cite{Howard2008, Eyles2009, Kano2008, Keller2007}. LASP has
heritage with readout electronics for a CCD closely related to the
CCD42-40 that are successfully operating on-orbit on the NASA
\textit{CUTE} mission\cite{Nell2021}.

Alongside the assembly of the CCD, we will collaborate with JPL to
develop a solar-blind AR coating for application to the
CCD\cite{Hennessy2015} to reject sensitivity across the visible
bandpass. This coating will be applied to another CCD42-40 with
enhanced passivation from Teledyne e2v, and this CCD will be
integrated into the F160M channel prior to a planned second flight of
FLUID. Performing this technology development on a device that already
has high TRL provides an expedited path for its use on larger class
missions with high reliability requirements.

\begin{figure}[!ht]
  \centering
  \includegraphics[width=\textwidth]{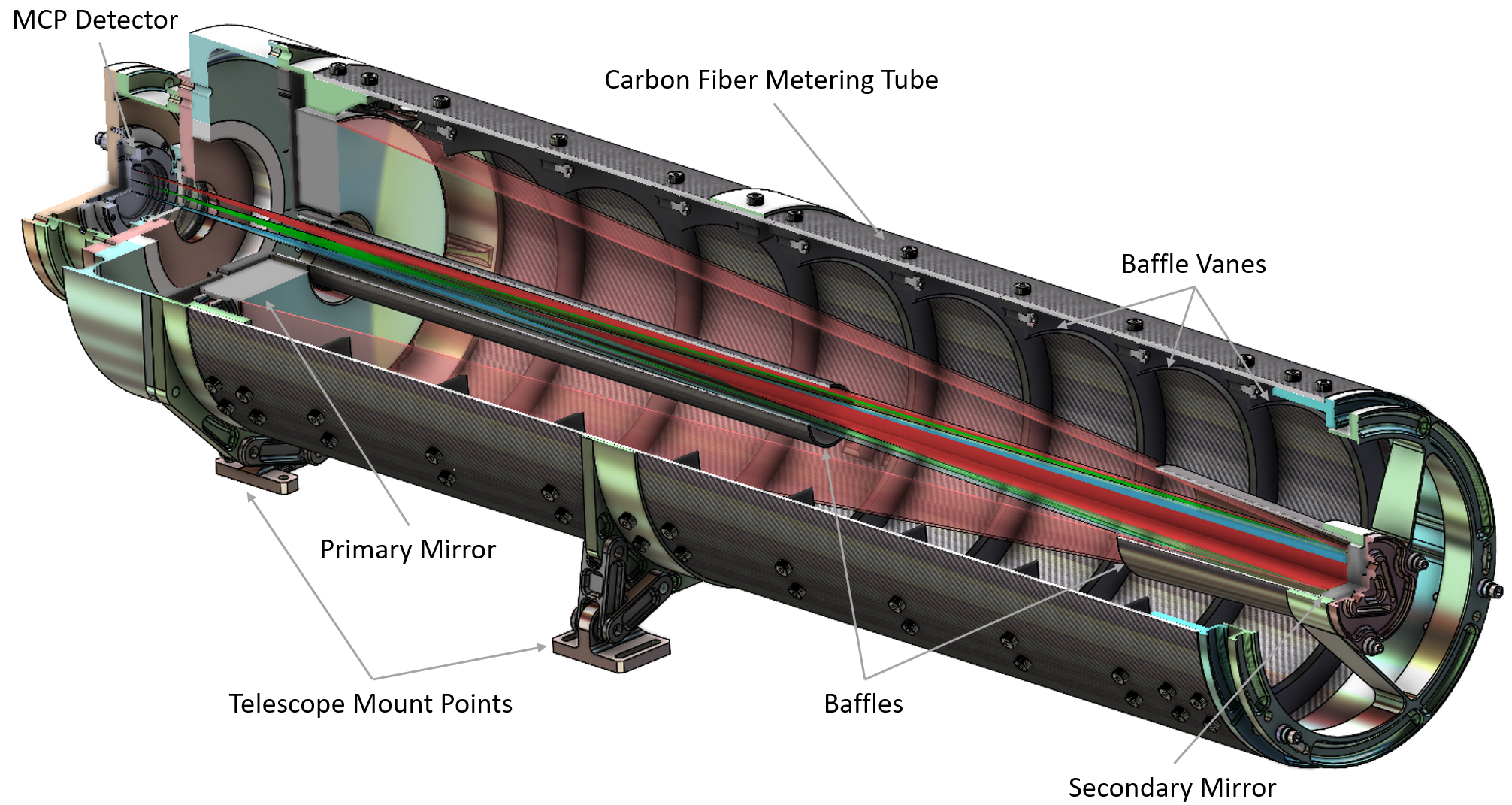}
  \caption{\label{fig:telescope}Cutaway CAD rendering of a single
    FLUID telescope. Light paths are shown in blue, green, and red
    corresponding to light falling on the center, midpoint, and edge
    of the detector respectively.}
\end{figure}

Multi-band imaging is used in other fields as well and the FLUID
instrument concept leverages aspects of the SDO-AIA
design\cite{Lemen2012}. Typically primary mirrors for astrophysical
missions are larger due to the desire to maximize signal-to-noise
ratios (SNR) for observations of faint targets. The multilayer coating
process used for the FLUID optics is currently limited to $\sim$150 mm
in diameter due to mechanical limitations of the coating chamber. The
sounding rocket envelope limits the largest possible optic diameter to
$\sim$500 \si{\mm}. While it is possible to build a large single
telescope with four integrated reflective filter imaging
channels, there are a variety of complications that make this
concept less desirable than independent optical channels. The
additional optics required to split and reimage the field after
filtering add efficiency losses and the system would be more
complicated to align and characterize. The independent channel design
allow each channel to be assembled, aligned, and characterized
independently and each channel has a mounting system with adjustment
to allow for coalignment of all the channels when they are mounted to
the sounding rocket optical bench (See Figures \ref{fig:telescope} and
\ref{fig:instrument}).

\subsection{Optical Mounts}
  
The optical mount for the primary mirror is based on the design of the
Interface Region Imaging Spectrograph (IRIS) primary mirror
mount\cite{Hertz2012, Podgorski2012a}, a design also adopted on the
Hi-C sounding rocket mission\cite{Podgorski2012b}. The flexures and
bond pads have been modified for the FLUID primary mirrors based on
anticipated launch loads and historical data from previous sounding
rocket missions. The use of tangential flexures minimizes the diameter
of the cell to allow all four channels to fit within the sounding
rocket envelope. We have also employed a cell with Maxwell kinematic
coupling points\cite{Slocum2010} to attach the primary mirror cell to
the optical metering tube assembly. The primary mirror cell can be
adjusted in tip and tilt to align the mirror on the optomechanical
axis of the system. The FLUID secondary mirror is sufficiently small
that it is possible to use a simpler ``pedestal'' style mount with a
single bond on the back surface of the optic. The secondary mirror
cell has integrated cutout style flexures to reduce the amount of
stress on the optic. Tip, tilt, and piston adjustment of the secondary
mirror is accomplished by moving the secondary spider structure which
is also mounted to the optical metering tube using a Maxwell kinematic
coupling. Once optimal alignment is achieved, the three mount points
are measured and the mounting hardware is trimmed to set the desired
distances. Fine adjustments are made with shims if necessary. This
method is used to ensure minimal possible movement or settling under
the strenuous vibration environment of sounding rocket flights.

\subsection{Instrument Resolution}

Final instrument imaging resolution is driven by three major
components: the telescope PSF, the detector response function, and the
attitude control system (ACS) jitter. The final observed resolution in
the instrument is the convolution of each of these
components\cite{Anderson2000}. The observation-driven specification
for the resolution of each channel is a PSF with \ang{;;4.2} full
width at half maximum (FWHM). The \ang{;;4.2} angular resolution
requirement corresponds to a spatial scale of 100 pc for the furthest
FLUID target. This spatial scale is selected to be compatible with
high-redshift scales targeted by future large UVOIR
missions\cite{Luvoir_2019} and to enable comparison with high-redshift
images obtained by JWST\cite{Rigby_2023}. The component and total
resolution estimates are shown in Table \ref{tab:resol}. Currently we
estimate a telescope PSF of \ang{;;1.2} FWHM through analysis using
Zemax OpticStudio software based on fabrication tolerances and
expected mechanical alignment tolerances. Due to the nature of the
position detection in XDL and XS anode MCP readouts, rather than
attempting to quantify the response function as the size of a pixel it
is advantageous to instead describe the position as a ``resolution
element'' (resol) since there are no physical pixels in the MCP
readout\cite{Vallerga2000, Vallerga2001}. The resolution element
describes the limiting imaging performance of the MCP detector as a
system incorporating all elements in the MCP detector that can affect
imaging resolution. Each type of detector has a different response
function. The XDL MCP detectors for F110M and F140M have
specifications of 30 \si{\um} FWHM resols. The F180M XS MCP detector
has a specification of 20 \si{\um} FWHM resols. The CCD42-40 has 13.5
\si{\um} square pixels. These resols can be described with a Gaussian
shape and thus we use a single number such as the FWHM of the resol to
describe the response function of XDL and XS anode MCP
readouts. Finally, the jitter of the ACS during observations must also
be considered. Using vehicle performance reports from past sounding
rocket flights we estimate a predicted ACS jitter FWHM of $<$
\ang{;;0.5}. Each of these components is mapped to angular space in
Table \ref{tab:resol} using the instrument plate scale (see Table
\ref{tab:instrument}) to directly compare and convolve magnitudes of
the different components. Ultimately, after considering all
contributions to the final resolution of FLUID, we predict that all
channels fall within the specification of \ang{;;4.2} FWHM with $>$
100\% margin.

\begin{table}[!ht]
  \centering
  \caption{FLUID In-band Resolution Budget}
  \label{tab:resol}
  \begin{threeparttable}
    \begin{tabular}{l l l l l l}
      \hline
      Channel & Telescope & Detector & ACS & Total (FWHM) & Margin\tnote{a} \\
      \hline
      F110M & \ang{;;0.8}\tnote{b} & \ang{;;1.4} & \ang{;;0.5} & $\le$\ang{;;2.0} & 110\%\\
      F140M & \ang{;;1.2}\tnote{c} & \ang{;;1.4} & \ang{;;0.5} & \ang{;;1.9} & 121\%\\
      F160M & \ang{;;1.2}\tnote{c} & \ang{;;0.7} & \ang{;;0.5} & \ang{;;1.4} & 200\%\\
      F180M & \ang{;;1.2}\tnote{c} & \ang{;;1.0} & \ang{;;0.5} & \ang{;;1.6} & 162\%\\
      \hline
    \end{tabular}
    \begin{tablenotes}[para, flushleft]
    \item[a] Note: the FLUID PSF requirement is \ang{;;4.2} FWHM.
    \item[b] Empirically characterized upper limit at visible wavelengths.
    \item[c] Predicted performance.
    \end{tablenotes}
    \end{threeparttable}
\end{table}

\subsection{Optomechanical Structural Analysis}

\begin{figure}[!ht]
  \centering
  \includegraphics[width=\textwidth]{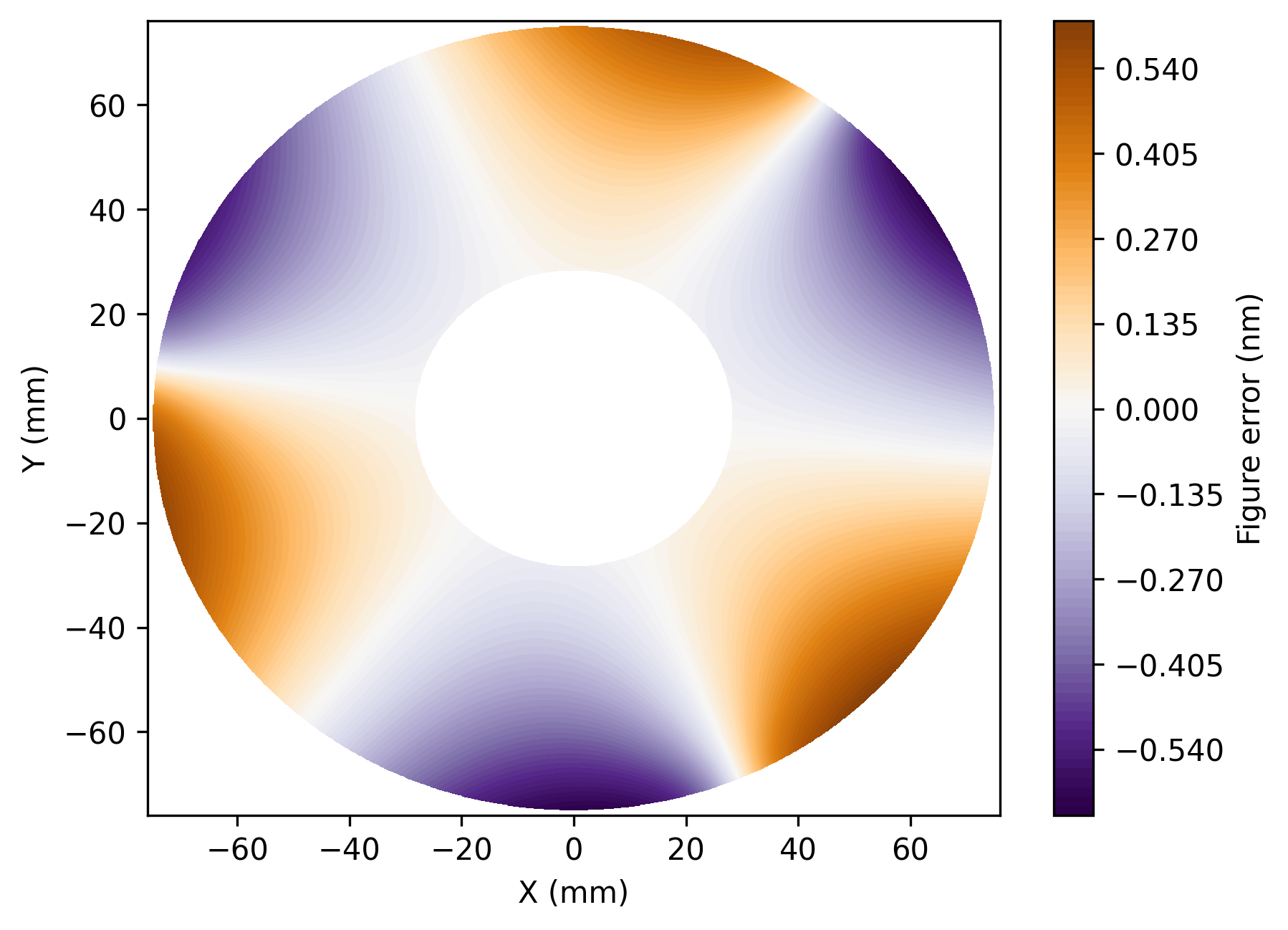}
  \caption{\label{fig:fea}Finite element analysis prediction of
    primary mirror figure error induced from the mechanical mount. The
    peak-to-valley value is 1.24 \si{\nm} with an RMS of 0.25
    \si{\nm}.}
\end{figure}

\begin{figure}[!ht]
  \centering
  \includegraphics[width=\textwidth]{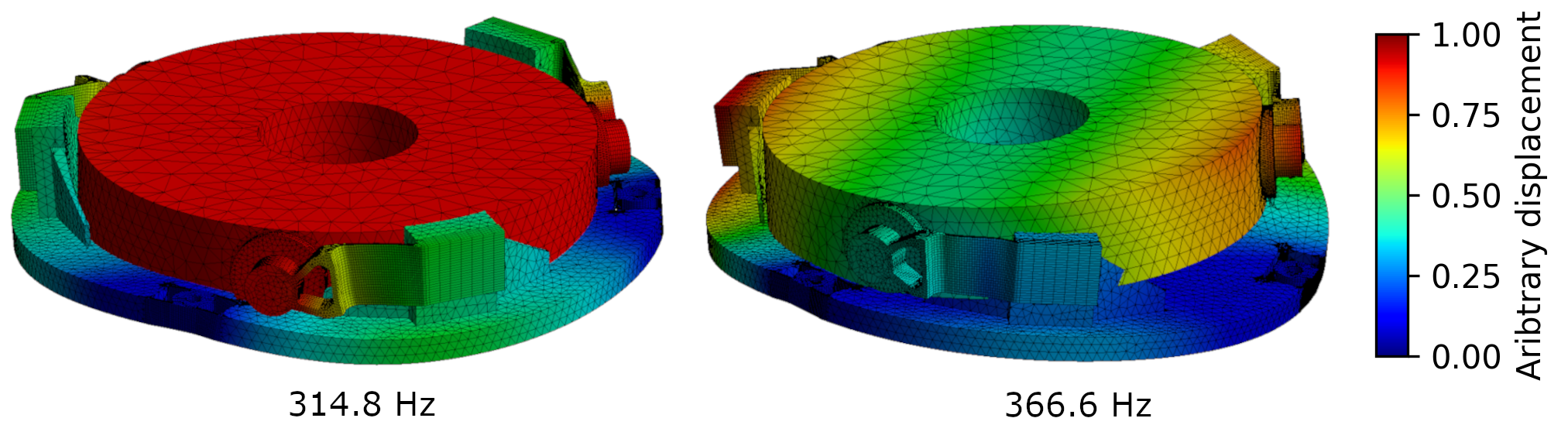}
  \caption{\label{fig:modal}Results of modal analysis of the FLUID
    primary mirror assembly. The first mode occurs at 314.8 \si{\Hz}
    (left) and the second mode occurs at 366.6 \si{\Hz}
    (right). Displacements generated by modal analysis (shown in the
    colormap) are qualitative and are thus shown on an arbitrary
    scale.}
\end{figure}

Structural analysis was performed on the primary mirror assembly to
verify that the design would survive the sounding rocket vibration
environment. A finite element model was constructed with Ansys to
analyze the primary mirror assembly. Static analysis with torque
values to be used for flight shows that the primary mirror surface
deforms less than 1.3 \si{\nm} peak-to-valley, shown in Figure
\ref{fig:fea}, which is well within the figure specification (See
Figure \ref{fig:psd}) required to achieve the desired PSF. This model
is used in conjunction with the vibration levels presented in Table
6.3.4-1 of the NASA Sounding Rockets User
Handbook\cite{NASA_NSROC_handbook} to analyze the primary mirror
assembly. The first and second modes occur at 314.8 \si{\Hz} and 366.6
\si{\Hz} respectively and are shown in Figure \ref{fig:modal}. The
first mode is primarily along the optical axis of the system and the
second mode is mainly lateral. These are the most notable modes and
analysis shows that the primary mirror assembly will handle the
vibration environment with substantial margin. Both static and
vibration analysis show that each component in the primary mirror
assembly exceeds the factor of safety outlined in the NASA structural
design standard document NASA-STD-5001A. A full thermal analysis has
not yet been performed but past data demonstrate that the temperature
range experienced by the telescopes in flight will be small due to the
isolation from the aluminum structure provided by the carbon fiber
optical metering tubes. The thermal effect on the optical metering
tube is minimal and well within the tolerances of the optical
system. The thermal isolation provided by the optical metering tube
allows a stable thermal operating environment for the optics and
optical mounts.

\section{TECHNOLOGY DEVELOPMENT}
\label{sec:techdev}

FLUID aims to accomplish development and maturation of several
technologies critical for future missions. These technologies include
FUV and LUV reflective bandpass filters on large substrates, sealed
tube MCP detectors with bialkali photocathodes and cross strip (XS)
anodes, passivated CCD detectors to improve QE at UV wavelengths, and
solar-blind multilayer coatings for silicon detectors.

\subsection{FUV Band-selecting Filters}
\label{sec:fuv_filters}

FLUID bandpasses are defined by reflective multilayer coatings
developed in collaboration with the Consejo Superior de
Investigaciones Cient\'ificas (CSIC) Grupo de \'Optica de L\'aminas
Delgadas (GOLD)\footnote{\url{https://gold.io.csic.es/}}. NASA's
2022 Astrophysics Biennial Technology Report lists ``High-Throughput
Bandpass Selection for UV'' and ``Far-UV Imaging Bandpass Filters'' as
Tier 1 and Tier 3 technologies, respectively, to be developed for
future NASA missions\cite{NASAbreport2022}. Initially, coating
prescriptions are modeled to perform trade studies on total
throughput, out-of-band rejection, and effective filter width. Once an
optimal prescription is identified, test depositions are performed on
2'' \texttimes{} 2'' witness samples, which are measured across the
bandpass of interest to verify performance. This process is repeated
until the sample matches the modeled prescription to within typical
deposition process, material, and modeling tolerances. Finally,
each flight optic is coated in a separate run alongside a witness
sample. Witness samples are measured on-site at a facility described
in References \citenum{Larruquert2011} and \citenum{Larruquert2022}
prior to shipment to LASP. Each witness sample and optic is then
measured at LASP in the ``square tank'' facility (See Figure
\ref{fig:aeff}) to verify performance of the flight optics versus
their witness samples and quantify any changes that may have occcurred
during shipment\cite{France2016, Windt1986}. Flight optics are
measured at multiple points across their clear aperture to quantify
uniformity of the bandpass filter. After measurements the optics are
stored in active nitrogen purge boxes and the witness samples are
stored in controlled humidity chambers. Witness samples are then
regularly measured to quantify degradation at different relative
humidity levels. Details of this process and measurements of the
coatings on flight optics are described in Reference
\citenum{Farr2023}.

\subsection{Solar-blind FUV Detectors}

NASA's 2022 Astrophysical Biennial Technology Report also lists
``Large-Format, Low-Darkrate, High-Efficiency, Photon-Counting,
Solar-blind, Far- and Near-UV Detectors'' as a Tier 1
priority\cite{NASAbreport2022}. For the F180M channel, a sealed tube
MCP detector with a semi-transparent bialkali photocathode and an XS
anode has been baselined in collaboration with the Space Sciences
Laboratory at UC Berkeley. This detector exploits the reliability of
commercially available sealed tube technology (the Photonis Planacon)
but substitutes the typical commercial anode, which is only capable of
low resolution imaging, for an XS anode capable of high resolution
imaging. The semi-transparent bialkali photocathode applied to the
window of the sealed tube has a short wavelength cutoff of $\sim$160
\si{\nm} due to the transmissivity of the window material and will be
optimized to reduce sensitivity at visible
wavelengths\cite{Siegmund2023, Siegmund2020}. The MCPs in this
detector will be processed via atomic layer deposition to tune the
resistance and secondary emission characteristics. Ultimately, this
results in an MCP detector with good gain and stability
characteristics\cite{Ertley_2018, Ertley_2015}.

For the first planned flight of the FLUID F160M channel we have
baselined a passivated CCD42-40 with an AR coating expected to achieve
a quantum detection efficiency of 30 -- 40\% over the F160M
bandpass\cite{Heymes2020a}. We will thoroughly characterize this
device in the FUV at predicted flight operating temperatures. A
benefit of the CCD42-40 is the ability to operate the device at
Peltier temperatures such that temperatures under $-40$\si{\celsius}
result in a system dominated by read noise. For characterization
testing we typically achieve these temperatures using a thermoelectric
cooler\cite{Nell2021} and for flight we will leverage a cooling design
from the LASP SDO-EVE calibration sounding rocket\cite{Hock2012}. The
unique facilities at LASP allow measurements of reflectivity of such a
device to be made across the FUV bandpass. Facilities at LASP designed
for detector characterization in the FUV also allow QE to be measured
for the device and in-band flat fields to be genereated\cite{Nell2016,
  Nell2021}. This suite of fundamental measurements enables
calculations of quantum yield and charge collection efficiency to be
made in the FUV bandpass to thoroughly characterize current
performance capabilities.

In collaboration with JPL, we plan to coat a passivated CCD42-40 with
a solar-blind metal-dielectric coating prescription (Al/Al$_2$O$_3$)
that maintains good detector efficiency ($\sim$20\%) in the F160M
bandpass while holding sensitivity across the entire visible bandpass
to a low level ($<$4\%)\cite{Hennessy2015}. Despite a drop in in-band
performance, it is the significant out-of-band suppression that is
critical for improving SNR in FUV observations with a CCD. CCDs
inherently have high QE (40 -- 65\%) across the full visible bandpass
due to the optical properties of silicon. Reducing the out-of-band
sensitivity by an order of magnitude significantly reduces background
noise from sources such as scatter, other objects in the FOV, and
out-of-band flux from target objects. This technology is vital for
improving SNR across the FUV bandpass where silicon is inherently less
sensitive. The coating performance will be characterized at LASP
through measurements of reflectivity and QE across the FUV and visible
bandpasses.

\section{ASSEMBLY}

\begin{figure}[!ht]
  \centering
  \includegraphics[width=\textwidth]{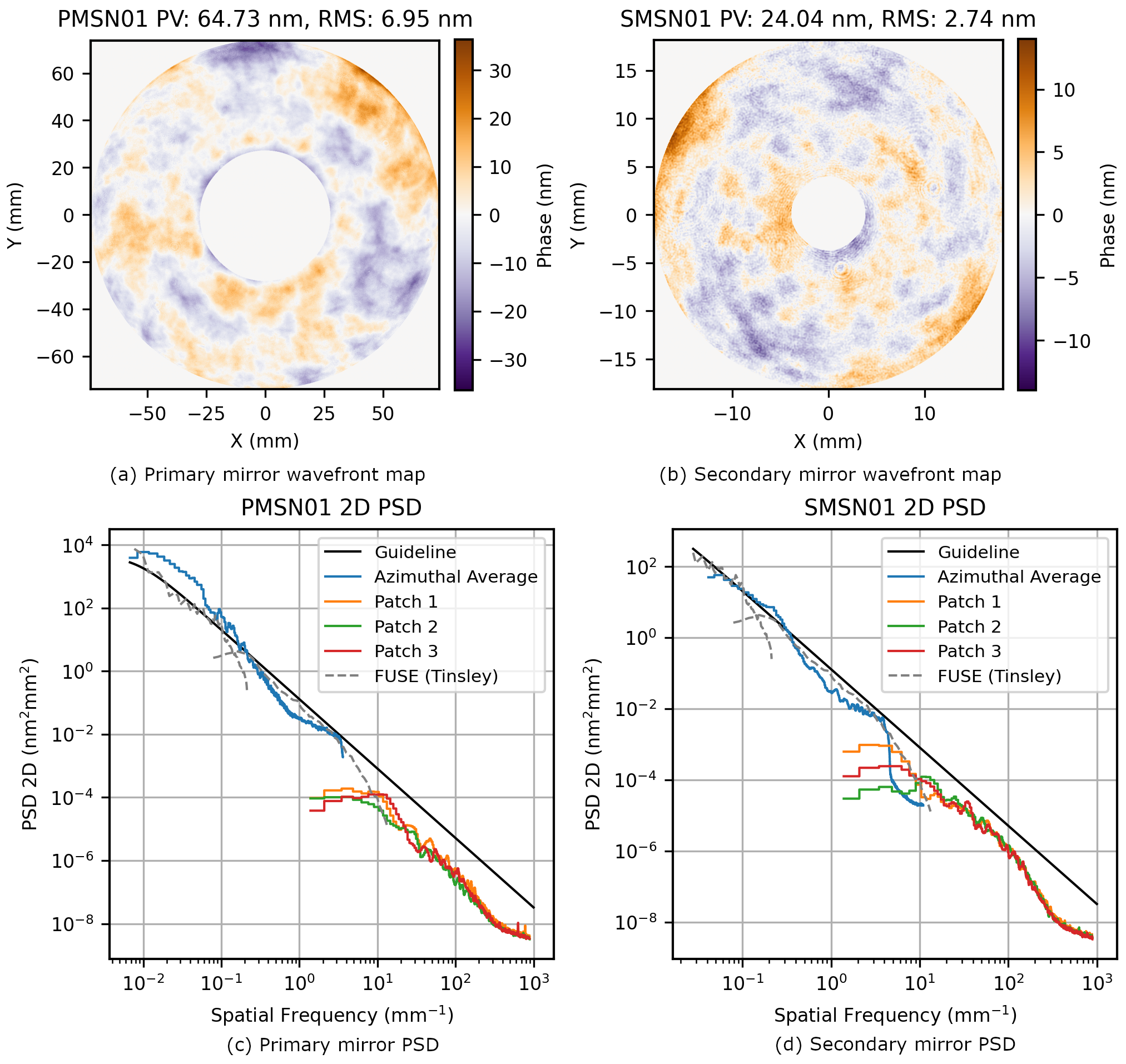}
  \caption{\label{fig:psd}Metrology data for the FLUID F110M primary
    and secondary mirrors. Wavefront maps for each mirror are shown in
    panels (a) and (b). Power spectral density data is shown in panels
    (c) and (d). Blue lines show azimuthally averaged 2D PSD data for
    the FLUID F110M mirrors. PSD data for higher spatial frequencies
    is shown for a series of small patches (sized 0.36 \texttimes{}
    0.27 \si{\mm}). {\em FUSE} PSD data from Reference
    \citenum{Ohl2000} is shown for comparison.}
\end{figure}

\begin{figure}[!ht]
  \centering
  \includegraphics[width=\textwidth]{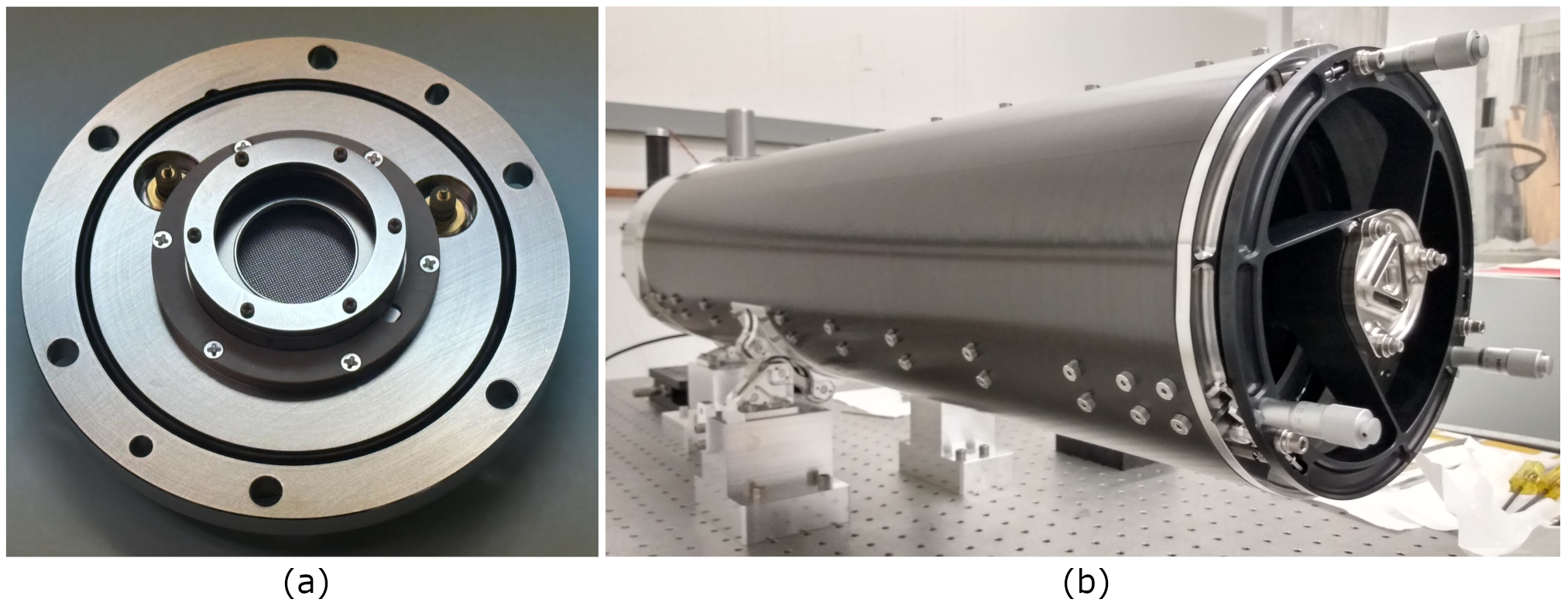}
  \caption{\label{fig:detector}(a) A FLUID open face XDL MCP
    detector. (b) The FLUID F110M telescope in the optical alignment
    setup.}
\end{figure}

The optical substrate material of the telescope optics is Schott
N-BK7, which is chosen to reduce stress in the multilayer coatings by
attempting to thermally match to the fluoride materials used in the
multilayer coating process\cite{LopezReyes2021,
  LopezReyes2022}. Metrology details for the F110M primary and
secondary mirrors prior to coating are shown in Figure
\ref{fig:psd}. The guideline power spectral density (PSD) shown in
Figure \ref{fig:psd} was designed to meet the projected final
telescope PSF of $\leq$ \ang{;;1.2} FWHM. The optical metering tube
and associated baffle tubes were acquired from Ability
Composites. Internal baffle tubes are bonded directly to the optical
cells to be well coaligned with the optical path to avoid
vignetting. The optical metering tube has been cleaned and vacuum
baked to verify cleanliness and low outgassing prior to having baffle
vanes installed (see Figure \ref{fig:detector}).

\begin{figure}[!ht]
  \centering
  \includegraphics[width=\textwidth]{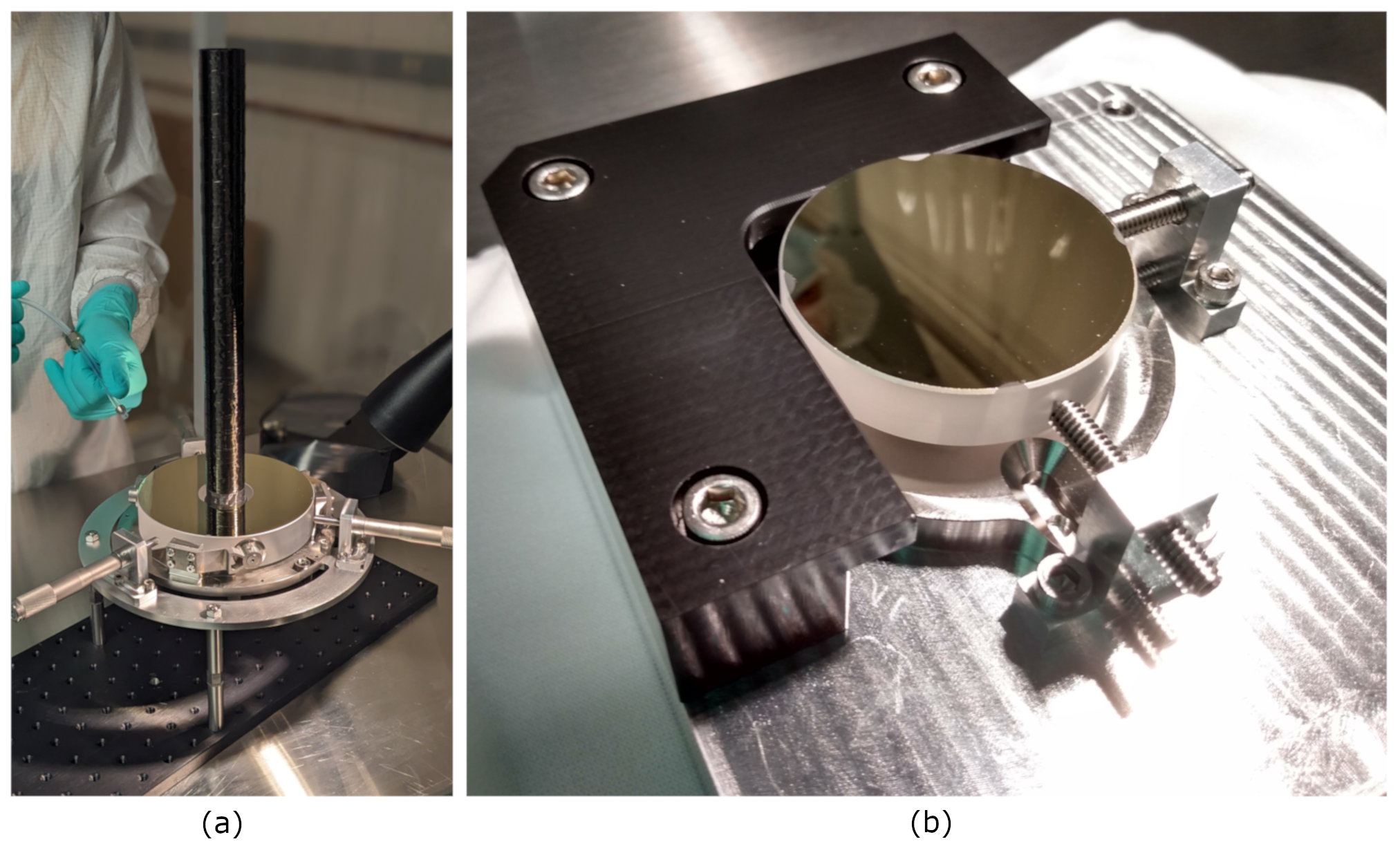}
  \caption{\label{fig:primary}(a) The F110M primary mirror bonded in
    place to the optical cell including central baffle. (b) The FLUID
    F110M secondary mirror in its bonding fixture.}
\end{figure}

The F110M and F140M telescopes are currently being assembled and all
telescope optics for F110M and F140M are coated, characterized, and in
nitrogen purged storage at LASP. An image of the F110M primary mirror
after being bonded to its cell is shown in Figure
\ref{fig:primary}. Prior to assembly all optics were characterized in
the LASP ``square tank'' facility and compared to their witness
samples. Results from this calibration run and additional aging
studies are discussed in detail in Reference
\citenum{Farr2023}. Bonding of the F110M secondary optic has been
completed and final telescope assembly and alignment occurred in
September through November 2023. Following the buildup of F110M the
same process will be repeated for F140M. The F110M and F140M MCP
detectors are currently undergoing optimization at Sensor Sciences LLC
and will be delivered in February 2024. The F110M detector assembly is
shown in Figure \ref{fig:detector}.

\section{TELESCOPE OPTICAL ALIGNMENT}

\subsection{Alignment Method}

Telescope alignment is accomplished in two steps. First, the primary
mirror cell is adjusted in tip/tilt to coalign the optical axis of the
primary mirror with the optical metering tube. To do this, the
telescope is mounted to an optical bench with the secondary mirror not
installed, a bright point source that exits a pinhole is placed on the
telescope optomechanical axis at a distance of the radius of curvature
of the primary mirror. The primary mirror cell is adjusted in tip and
tilt via shim under the kinematic mount points until the return point
is coaligned with the exit pinhole of the source. The primary cell is
then locked in place and the secondary mirror is installed. A double
pass system is used to align the secondary mirror of the telescope. A
10 inch flat mirror is placed away from the aperture and a point
source microscope (PSM) is placed at the focal plane. The PSM is
similar in design to those shown in References \citenum{Steel1983},
\citenum{Kuhn2016}, and \citenum{Parks2005} and was designed and built
at LASP for optical alignments. The light source used in conjunction
with the PSM has a wavelength of 530 \si{nm} and an adjustable source
diameter. The secondary mirror is adjusted in tip, tilt, and piston
until the return PSF observed in the PSM is symmetric with a profile
that is consistent with optical modeling including fabrication and
mounting tolerances of the system. Initially the secondary mirror is
moved in tip, tilt, and piston with three micrometers. This allows
verification of the telescope performance and in this state shim sizes
for fixing the secondary mirror into place are estimated. Final
secondary mirror adjustments are made by removing the micrometers and
installing shims under the kinematic mount points to adjust the
secondary mirror in tip, tilt, and piston. Secondary mirror
adjustments are decoupled to simplify the process, first, the tip and
tilt are determined and those adjustments are made until the return
spot is centered on the focal plane to within acceptable
tolerance. Next, piston adjustments are made to focus the telescope
onto the mechanical focal plane location. All optical mounts have been
torqued into place using torque values suitable for flight. Final
verification will be performed in-band in the LASP ``longtank''
facility. This facility is a large (30 inch inner diameter) vacuum
chamber with a 24 inch Newtonian collimator illuminated with a flowing
gas hollow cathode discharge lamp\cite{Cook1991}.

\subsection{Alignment Results}

\begin{figure}[!ht]
  \centering
  \includegraphics[width=\textwidth]{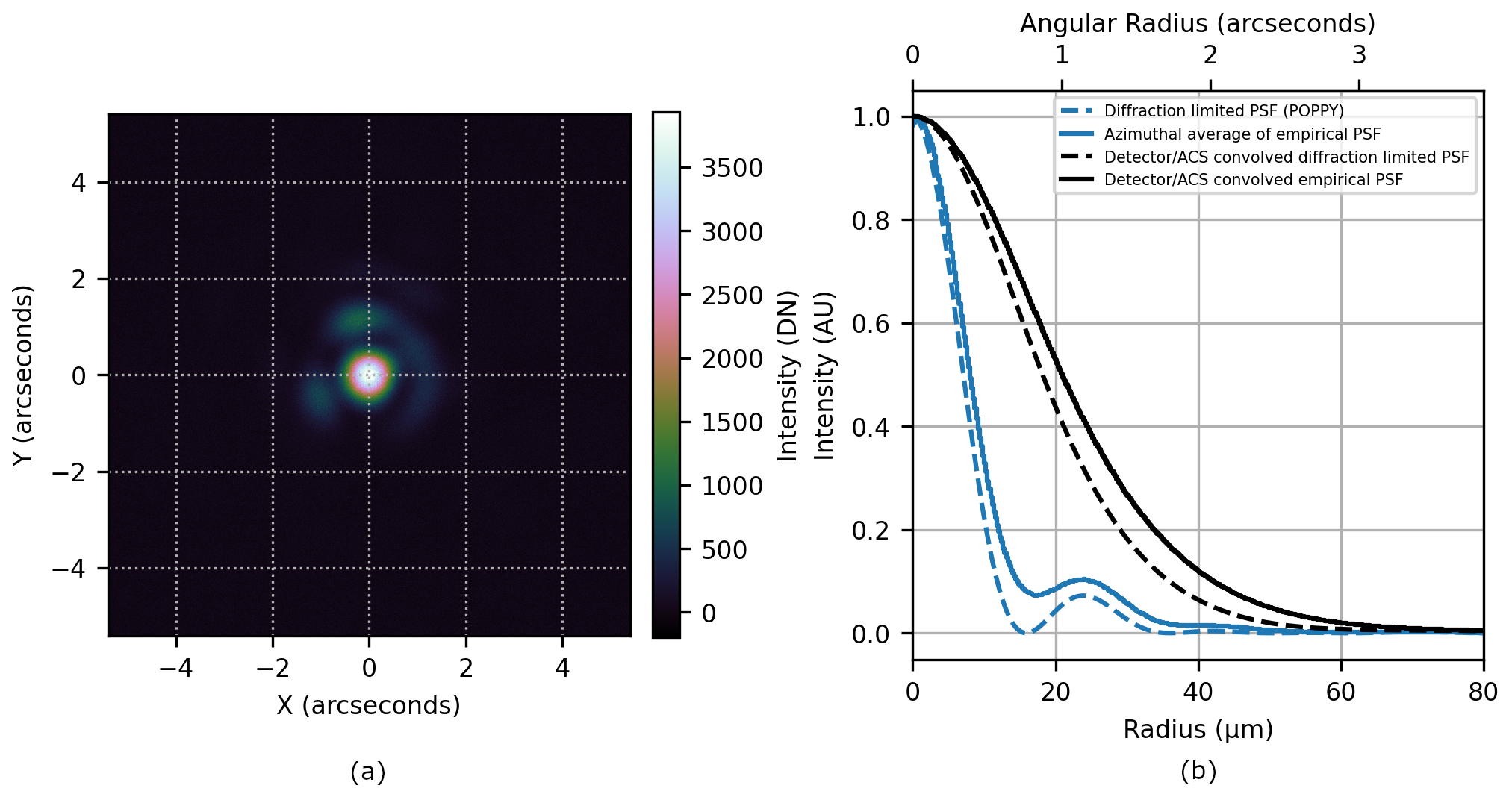}
  \caption{\label{fig:psf}(a) shows the FLUID F110M PSF as observed
    with the PSM using 530 \si{\nm} light. Diffraction features are
    clearly visible. Panel (b) shows the azimuthally averaged profile
    of the PSF shown in panel (a) along with the predicted performance
    after factoring in the detector resolution element and estimated
    ACS performance. For comparison we also show the theoretical
    diffraction limited performance.}
\end{figure}

The FLUID F110M telescope achieves nearly diffraction limited
performance in the visible bandpass. The measured and predicted PSF
are shown in Figure \ref{fig:psf}. It is critical to note that the PSF
displayed in Figure \ref{fig:psf} contains double the wavefront error
of the telescope itself due to the nature of the light path in a
double pass autocollimation test. However, because we do not currently
have a method to measure the wavefront error of this system, it is not
straightforward to remove excess contributions from the double pass
test results. Therefore, this result overestimates aberration in the
system. Additionally this telescope has been tested horizontally but
is designed to operate in free fall during a sounding rocket
flight. Finite element analysis of the gravity load on the primary
mirror flexures shows that some astigmatism will be introduced to the
primary mirror surface in the testing orientation under gravity
load. The source used to illuminate the telescope for this alignment
has a diameter of 10 \si{\um} and a ``tophat'' shape. Although the PSM
is capable of producing smaller source diameters this size is used to
provide sufficient intensity to produce a suitable SNR on the
sensor. A source diameter of $\sim$ 3.2 \si{\um} or smaller would be
required to properly resolve the PSF\cite{Smith1998}. Additionally
this PSF has been characterized with a 530 \si{\nm} source but because
the F110M central wavelength is 105 \si{\nm} we expect the results
from testing at 530 \si{\nm} to scale with wavelength when F110M is
illuminated with in-band light. Ultimately, considering all above
contributions to the measured PSF, we expect the results shown in
Figure \ref{fig:psf} to be an upper limit estimate on the in-band
performance of the F110M telescope. Figure \ref{fig:psf} shows modeled
contributions to the PSF from the detector resolution element and
jitter of the ACS in flight. Additionally, the theoretical diffraction
limited performance computed with the POPPY software\cite{Perrin2012}
is shown along with the detector and ACS convolved diffraction limited
performance as a comparison against empirical F110M performance. With
all contributions included we predict a PSF for F110M with an FWHM
$\le$ \ang{;;2.0}, resulting in a margin of 110\% for the FLUID
\ang{;;4.2} requirement.

\section{SUMMARY}

This paper has described the design and current state of the FLUID
instrument, a four channel multi-band LUV and FUV imaging system. We
have described the expected optical performance over the full bandpass
of 92 -- 193 nm and the technology required to achieve the desired
performance in each channel. The executed and planned technology
development of FUV multilayer coatings, sealed tube MCPs, passivated
CCDs, and solar-blind CCD multilayer coatings are summarized and their
important role in optimization of each of the FLUID channels is
discussed. We have presented the current state of the FLUID F110M and
F140M channels, which will continue to be assembled, tested, and
characterized through the summer of 2024. The optical alignment
results for F110M meet requirements with a significant margin and
demonstrate a robust design that will be repeated for the remaining
FLUID telescopes. The FLUID team is working hard to demonstrate the
instrument capability with the funding available currently and will
propose for funding though the 2023 ROSES/APRA call to complete the
FLUID optical system development, advance the optical and sensor
technology as described above, and realize the full potential of the
FLUID concept with observations of several galaxies across three
sounding rocket flights.


\subsection* {Code, Data, and Materials Availability} 

The data that support the findings of this paper are not publicly
available. They can be requested from the author at
\url{nicholas.nell@lasp.colorado.edu}. 

\acknowledgments 
 
The FLUID team would like to thank GOLD for their excellent coatings
and great attention to detail throughout the process of developing,
applying, and characterizing the multilayer coatings for FLUID. This
research made use of POPPY, an open-source optical propagation Python
package originally developed for the {\em James Webb Space Telescope}
project\cite{Perrin2012}. We thank Daniel Warren of CIRES for efforts
on fabricating optical mounting hardware. This work was supported by
NASA grant 80NSSC21K2016 to the University of Colorado.

\FloatBarrier

\bibliography{fluidbib} 
\bibliographystyle{spiejour} 

\end{document}